\documentclass[twocolumn,english,showpacs,prb,floatfix]{revtex4}%
\usepackage{amsfonts}
\usepackage{amsmath}
\usepackage{amssymb}
\usepackage{graphicx}%
\begin{document}
\title{First-Principles Calculation of Be(0001) Thin Films: Quantum Size Effect and
Adsorption of Atomic Hydrogen }
\author{Ping Zhang, Hong-Zhou Song}
\affiliation{Institute of Applied Physics and Computational Mathematics, P.O. Box 8009,
Beijing 100088, P.R. China}

\begin{abstract}
We have carried out first-principles calculations of Be(0001) thin films to
study the oscillatory quantum size effects exhibited in the surface energy,
work function, and binding energy of the atomic hydrogen monolayer adsorption.
The prominent enhancement of the surface density of states at the Fermi level
makes Be(0001) thin films more metallic compared to the crystalline Be. As a
result, the calculated energetics of Be films and the properties of atomic H
adsorption onto Be(0001) surface are featured by a quantum oscillatory
behavior. Furthermore, The prominent change in the Be(0001) surface electronic
structure by the atomic hydrogen adsorption has also been shown.

\end{abstract}
\pacs{73.61.-r, 73.20.At, 73.21.Ac,}
\maketitle

\section{Introduction}

When the thickness of thin films approaches the nanoscale, the oscillatory
quantum size effects (QSE) associated with electronic confinement and
interference will occur\cite{Pag,Tes,Mey,Kaw} due to the splitting of the
energy-level spectrum into subbands normal to the plane of the films. It has
been shown that a change in film thickness by just one atomic layer can result
in property variations on the order of $1/N$, where $N$ is the thickness of
the film in terms of monolayers (ML). The oscillatory QSE have long been
clearly observed in ultrathin metal overlayers on metal
substrates\cite{Miller}. Recent systematic experimental and theoretical
investigation of the QSE has mainly been focused to Pb films deposited on
Si(111)\cite{Chen1,Yeh,Su,Chen2,Chiang1,Chiang2,Chiang3,Dil,Xue1,Chiang4,Chiang5,Chiang6,Xue2,Zhang,Silva,Ogando}
or Cu(111)\cite{Mat} substrates.

In this paper we present a detailed first-principles study of the electronic
structure and adsorption energetics of Be free standing films. The QSE in
other simple-metal films such as Al\cite{Fei,Kie,Cir}, Mg\cite{Fei},
Li\cite{Boe}, Rh\cite{Fei}, and Pb\cite{Saa,Wei} have been reported in
previous references. Although Be is also a simple metal, its electronic
structure is profound compared to its close neighbors in the Periodic Table.
In particular, the electronic properties of bulk Be are very distinct from
those in Be surface. For instance, the electron density of states (DOS) in
bulk Be has a very deep minimum at Fermi level ($E_{F}$), making this material
nearly semiconducting\cite{Pap,Yu} while in the Be(0001) and Be(10\={1}0)
surfaces the DOS at $E_{F}$ was found to be larger by a factor of 4-5 that
makes these surfaces two-dimensional free-electron-like metals\cite{Chu}.
Compared to Mg and Li which have been previously studied\cite{Fei,Boe} in QSE,
the bonding energy of Be is surprisingly strong (cohesive energy Be: 3.32
eV/atom, Mg: 1.51 eV/atom, Li: 1.63 eV/atom\cite{Kittle}). Also the Debye
temperature is very high (Be: 1440 K, Mg: 400 K, Li: 344 K\cite{Ame}). Thus
quantum effects should be more pronounced in determing low-temperature
structural properties\cite{Laz}.

This paper is organized as follows: In Sec. II, the \textit{ab initio} based
method and computational details is outlined. In Sec. III, the surface
properties of the Be(0001) films, including the electronic structure, surface
energy, work function, and interlayer relaxation, as a function of the
thickness of the films, are presented and discussed. Also the properties of
adsorption of atomic hydrogen monolayer onto Be(0001) surface is discussed in
detail by presenting the sensitivity of the adsorption energy and local DOS to
the thickness of the Be(0001) films. Finally, Sec. IV contains a summary of
the work and our conclusion.

\section{Computational method}

The calculations were carried out using the Vienna \textit{ab initio}
simulation package\cite{Vasp} based on density-functional theory with
ultrasoft pseudopotentials\cite{Vand} and plane waves. In the present film
calculations, free-standing Be(0001) films in periodic slab geometries were
employed. The periodic slabs are separated by a vacuum region equal to 20
\AA . In all the calculations below, a surface ($1\times1$) was employed for
the supercell slab. The Brillouin-zone integration was performed using
Monkhorse-Pack scheme\cite{Pack} with a $11\times11\times1$ $k$-point grid,
and the plane-wave energy cutoff was set $300$ eV. Furthermore, the
generalized gradient approximation (GGA) with PW-91 exchange-correlation
potential has been employed with all atomic configurations fully relaxed.
First the total energy of the bulk hcp Be was calculated to obtain the bulk
lattice constants. The calculated $a$- and $c$-lattice parameters are $2.272$
\AA and $3.544$ \AA , comparable with experimental\cite{Amo} values of $2.285$
\AA and 3.585 \AA , respectively. The use of larger $k$-point meshes did not
alter these values significantly. A Fermi broadening of 0.1 eV was chosen to
smear the occupation of the bands around $E_{F}$ by a finite-$T$ Fermi
function and extrapolating to $T=0$ K.

\section{Results and discussion}

\subsection{Band structure}%

\begin{figure}[tbp]
\begin{center}
\includegraphics[width=1.0\linewidth]{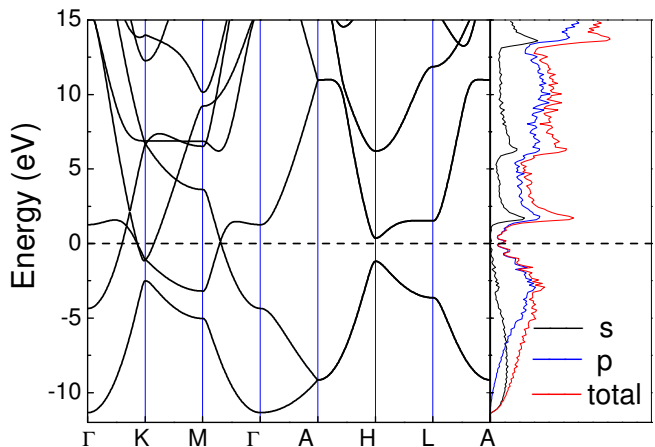}
\end{center}
\caption
{(Color online) GGA energy bands and density of electron states (right panel) of hcp bulk Be.
The dashed line denotes Fermi level.} \label{fig1}
\end{figure}%
We first studied the properties of electronic structures of Be(0001) films. As
a first step, we present in Fig. 1 the band structure and the DOS of bulk hcp
Be. One can see that the density of electronic states for bulk Be resembles
somewhat that of a semiconductor since it has a minimum at the Fermi energy.
This makes Be different from its close neighbors, such as Mg, whose band
property is nearly free-electron like. The bulk Be bands display a direct gap
in a large range of the Brillouin zone, unlike Mg. Mg has a filled state at
$\Gamma$ with energy $\simeq1.3$ eV\cite{Laz}, while the corresponding Be
state is above the Fermi energy and its band is nearly flat. This band is the
source of both the low density of states near the Fermi energy and the high
peak above the Fermi energy. Although the electronic configuration of
elemental Be is $1s^{2}2s^{2}$, one can see from Fig. 1 that the $p$-orbital
component in bulk Be plays a main role around $E_{F}$. The bonding properties
of hcp Be is anisotropic, which can be seen by the fact that the $c/a$ ratio
(1.56) is one of the most contracted for hcp metals (for Mg, $c/a\simeq1.62$).
Thus out-of-plane neighbors have shorter bonds than in-plane neighbors.
Another evidence (not shown here) of this anisotropy is that the contribution
of $p_{x}$ and $p_{y}$ orbitals to the DOS is different from that of $p_{z}$
component\cite{Cohen}.%

\begin{figure}[tbp]
\begin{center}
\includegraphics[width=1.0\linewidth]{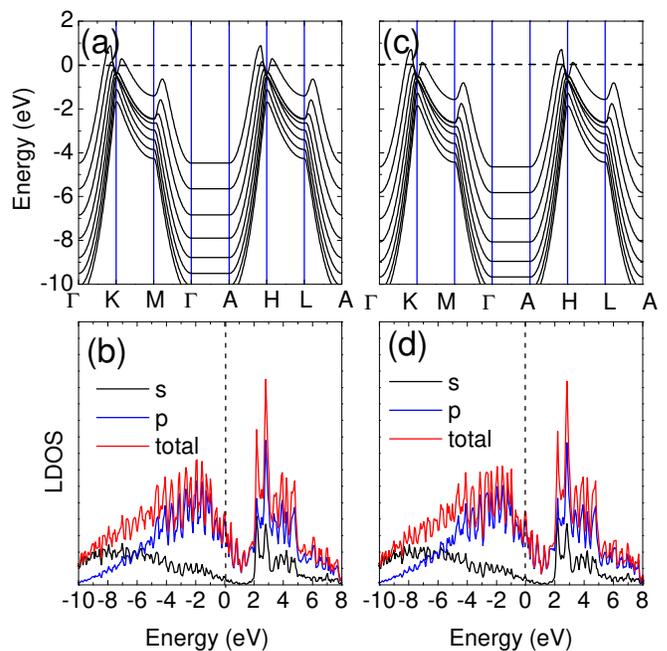}
\end{center}
\caption
{(Color online) Surface band structures and density of states of a Be(0001) slab with 9 Be monolayers.
The left and right panels are for the unrelaxed and relaxed geometries, respectively.}
\label{fig2}
\end{figure}%
The above-mentioned "semiconducting" metallic picture for bulk Be are totally
changed in the case of Be(0001) surface. To illustrate this change, the
electronic structure properties of Be(0001) film of 10 layers are shown in
Fig. 2, wherein the left panels give the results for the unrelaxed structure,
while the right panels show the relaxed results. For a film with the thickness
of 10 layers, it shows in Fig. 2 no prominent difference in electronic
structures between relaxed and unrelaxed slabs. Compared to the bulk results
shown in Fig. 1, it reveals in Fig. 2 that (i) the Fermi level in the Be(0001)
film shifts down towards the broad peak, resulting in a prominent enhancement
of the DOS at $E_{F}$ with respect to the bulk case. This phenomenon of high
density of surface states at $E_{F}$ was previously reported\cite{Chu}, and
was also noticed in other metal films such as W(110) and Mo(110)\cite{Rot};
(ii) the band structure of the Be(0001) slab is characterized by a series of
subbands which can be well fit by $E_{n}+\hbar^{2}(k_{x}^{2}+k_{y}%
^{2})/2m^{\ast}$ with the effective mass $m^{\ast}$. This clearly demonstrates
the quasi-2D free-electron character in the Be(0001) thin \ film, which is
contrary to the bulk Be.

\subsection{Energetics}

Figure 3(a) shows the total energy per monolayer $E(N)/N$ as a function of the
number $N$ of layers in the slab. The atoms in the slabs have been fully
relaxed during the calculations. One can see from Fig. 3(a) that with
increasing the thickness, $E(N)/N$ gradually approaches a constant value which
in the limit is equal to the energy per atom in the bulk Be.%

\begin{figure}[tbp]
\begin{center}
\includegraphics[width=1.0\linewidth]{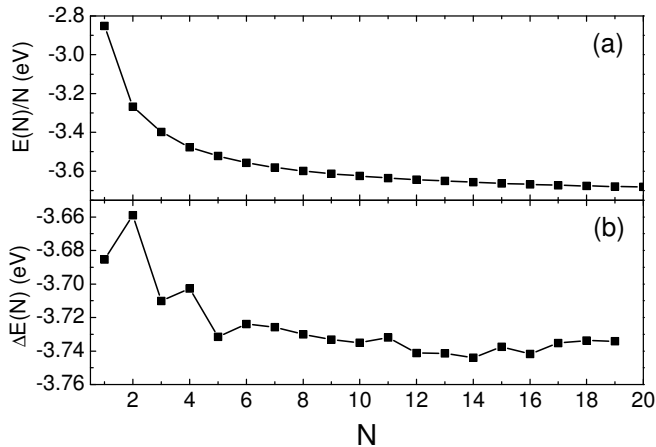}
\end{center}
\caption{(a) Monolayer energies $E(N)/N$ for fully relaxed Be(0001) slabs as
a function of the number of Be monolayers $N$; (b)
corresponding energy differences .} \label{fig3}
\end{figure}%
A quantity more suitably tailored to the QSE is the energy difference $\Delta
E(N)=E(N)-E(N-1)$. This quantity is shown in Fig. 3(b), which reveals damped
oscillations as a function of $N$. These oscillations arise from the
occupation of electronic levels close to Fermi surface\cite{Sch}, which are
$p$-like as shown in Fig. 2.%

\begin{figure}[tbp]
\begin{center}
\includegraphics[width=1.0\linewidth]{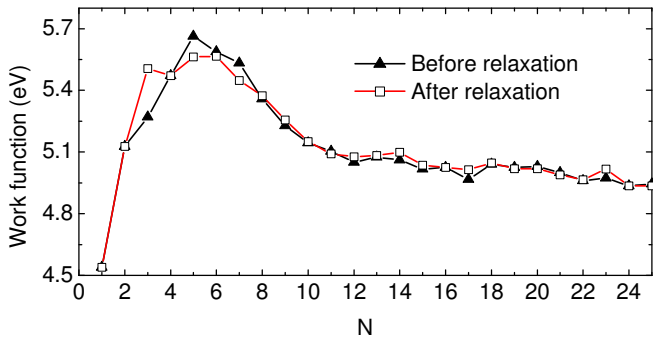}
\end{center}
\caption
{Work function of Be(0001) thin film as a function of the number of Be monolayers in
the slab.}
\label{fig3}
\end{figure}%
Figure 4 shows the work function of the Be(0001) thin films as a function of
the number of Be layers in the slab for the unrelaxed and the relaxed
geometries. One can see that the large influence of the relaxation on the work
function occurs in the range of $N\sim3$-$7$. For more larger values of $N$,
the overall effect of geometry relaxation on the work function trend is small.
Note that unlike the other metals such as Pb(111), in which the work function
oscillates periodically even in the ultra-thin film limit, the work function
of Be(0001) first curves up at small values of $N$ ($\sim3$), then lowers down
towards the bulk value in an oscillatory way. The peculiar behavior of work
function in Fig. 4 implies the more complicated thin-film properties of Be
compared to the other simple $s$-$p$ metals. The similar behavior also occurs
in the adsorption energy of atomic hydrogen onto Be(0001) surface (see below).

\subsection{Interlayer relaxation}

\begin{table}[th]
\caption{Interlayer relaxations given in percent, $\Delta d_{i,i+1}$, of
Be(0001) as a function of the number $N$ of layers in the slab.}
\begin{tabular}
[c]{ccccccc}\hline\hline
{$N$} & $\Delta d_{12}$ & $\Delta d_{23}$ & $\Delta d_{34}$ & $\Delta d_{45}$
& $\Delta d_{56}$ & $\Delta d_{67}$\\\hline
2 & +10.972 &  &  &  &  & \\
3 & +6.546 & +6.546 &  &  &  & \\
4 & +5.4 & +2.298 & +5.4 &  &  & \\
5 & +3.127 & +0.765 & +0.765 & +3.127 &  & \\
6 & +2.533 & -0.181 & +0.762 & -0.181 & +2.533 & \\
7 & +2.419 & -0.266 & +0.062 & +0.062 & -0.266 & +2.419\\
8 & +1.718 & -0.767 & +0.176 & -0.29 & +0.175 & -0.767\\
9 & +1.5 & -0.827 & -0.081 & +0.052 & +0.052 & -0.081\\
10 & +1.032 & -0.829 & -0.001 & -0.025 & +0.173 & -0.025\\
11 & +1.158 & -0.920 & -0.071 & +0.028 & +0.092 & +0.092\\
12 & +0.935 & -1.008 & -0.003 & -0.214 & +0.277 & -0.073\\
13 & +1.228 & -0.902 & -0.09 & -0.011 & +0.091 & +0.031\\
14 & +1.038 & -0.951 & -0.031 & -0.21 & +0.216 & -0.073\\
15 & +1.228 & -1.024 & +0.022 & -0.225 & +0.1789 & -0.018\\\hline\hline
\end{tabular}
\end{table}

The interlayer relaxation, $\Delta d_{i,i+1}$, are given in percent with
respect to the unrelaxed interlayer spacings, $d_{0}$, i.e., $\Delta
d_{i,i+1}=100(d_{i,i+1}-d_{0})/d_{0}$. $d_{i,i+1}$ is the interlayer distance
between two adjacent layers parallel to the surface calculated by total energy
minimization. $d_{0}=c/2$ is the bulk interlayer distance along (0001)
direction. As mentioned above, all layers in the slab were allowed to relax.
Obviously, the signs $+$ and $-$ of $\Delta d_{i,i+1}$ indicate expansion and
contraction of the interlayer spacings, respectively. The relaxation of
Be(0001) surface as a function of the number of layers is summarized in Table
I. Furthermore, the interlayer relaxations are also plotted in Fig. 5 as a
function of $N$ for clear illustration. One can see: (i) The three outmost
layers relax significantly from the bulk value, in agreement with experimental
observation\cite{Poh}. In the whole range of layers that we considered, the
topmost interlayer relaxation is always outward ($\Delta d_{12}>0$), with
$\Delta d_{12}$ starting from the largest value of +11\% for a slab with only
two monolayers, and approaches a final value of $\sim$1\% with increasing the
thickness of Be(0001) films. Note that the first interlayer separation on most
metal surfaces is contracted, Be(0001) is one of the few exceptions.
Considering that the calculation is given at zero temperature, our calculated
result of $\Delta d_{12}\simeq+1\%$ can be considered comparable with the
experimental observation of $\Delta d_{12}\simeq+3\%$ at T=110 K. (ii) When
the number of monolayers in the slab reaches $N\simeq5$, the interlayer
spacings begin to oscillate with a damped magnitude. Again, the oscillations
reflect QSE in Be(0001) thin films. After 25 monolayers, which is the maximal
layers considered here, the oscillations are still visible which indicates
that the semi-infinite surface limit is still not reached.%
\begin{figure}[tbp]
\begin{center}
\includegraphics[width=1.0\linewidth]{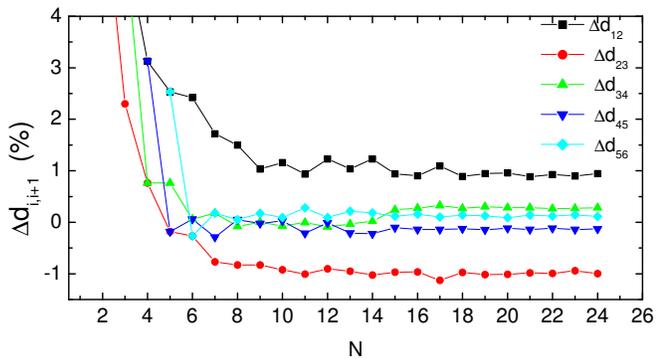}
\end{center}
\caption{(Color online) Relaxations of the Be(0001) surfaces as a function
of the number $N$ of Be monolayers in the slab} \label{fig5}
\end{figure}%

\subsection{Adsorption of atomic hydrogen: QSE of binding energy}

To further illustrate the physical properties influenced by finite size of the
thin films, in this section we focus our attention to the adsorption of atomic
hydrogen on Be(0001) thin films. To the best of our knowledge, the reflection
of QSE by the adsorption features has not been previously studied.%

\begin{figure}[tbp]
\begin{center}
\includegraphics[width=0.8\linewidth]{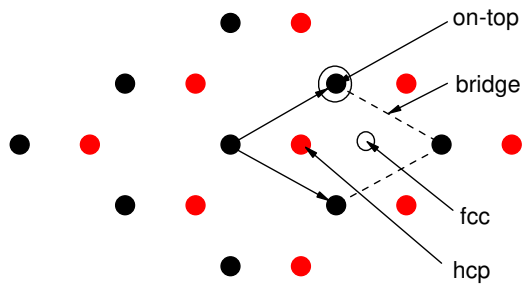}
\end{center}
\caption{(Color online) Adsorption sites for H adatoms on the
hcp Be(0001) surface. The indicated adsorption sites
are on-top, hcp, bridge, and fcc. The H adatoms and Be
atoms in the topmost surface layer are indicated by large
open circles and small black circles, respectively, while the
second-layer Be atoms are depicted as small red circles.} \label{fig6}
\end{figure}%
Before we study the hydrogen adsorption properties as a function of the
thickness of Be(0001) thin films, we need to determine the energetically
favourable adsorption site. Since the preference of adsorption site is not
sensitive to the thickness of the substrate, thus to look for this preference
of the adsorption site, it is sufficient to give a study on the slabs with
fixed thickness of the Be(0001) substrate, which at present is chosen to be 9
monolayers. We choose four most probable adsorption sites, namely, on-top,
bridge, fcc, and hcp sites, which are indicated in Fig. 6. The binding energy
is calculated using the following equation: Binding energy [atomic H]$=-$%
($E[$H/Be(0001)$]-E[$Be$(0001)]-2E[$H$]$)/2 where $E[$H/Be(0001)$]$ is the
total energy of a slab which consists of 9 layers of Be atoms and one H atom
on each side keeping inversion symmetry, $E[$Be$(0001)]$ is total energy of
the slab without H atoms, and $E[$H$]$ is total energy of a free H atom which
is put in a 10 \AA $\times$10 \AA $\times$10 \AA supercell. As a result, the
calculated binding energy of atomic H for different adsorption configurations
is 1.46 eV (on-top), 2.24 eV (fcc), 2.11 eV (hcp), and 2.33 eV (bridge),
showing a clear preference for bridge site adsorption. We note that this
bridge-site preference for H adsorption was also reported in Ref.[43] and in
other metal surfaces such as W(100)\cite{Bar}.%

\begin{figure}[tbp]
\begin{center}
\includegraphics[width=1.0\linewidth]{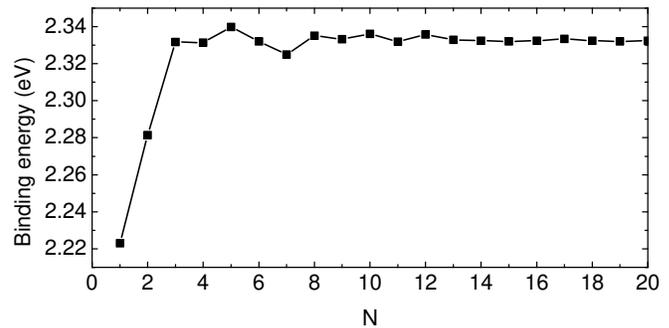}
\end{center}
\caption
{Calculated binding energy of H adatom as a function the number of Be(0001) monolayers
in the slab.}
\label{fig7}
\end{figure}%
After finding the preferred atomic H adsorption site (bridge site), we give a
series of calculations for the binding energy of the H adsorbate as a function
of the thickness of Be(0001) thin films. The results are summarized in Fig. 7.
One can see that the binding energy curves up at small film thickness,
followed by damped oscillations when increasing the Be monolayers in the slab.
Thus the binding energy of atomic H depends on the thickness of the quantum
films in an oscillatory way. In experiment this dependence can be observed by
investigating the dependence of H coverage on the monolayers of Be(0001) thin
films. Note that the dependence of H adatom binding energy on the thickness of
the Be(0001) films looks in analogy with the work function behavior shown in
Fig. 4. The cause of this similarity remains unclear at the present stage.%

\begin{figure}[tbp]
\begin{center}
\includegraphics[width=1.0\linewidth]{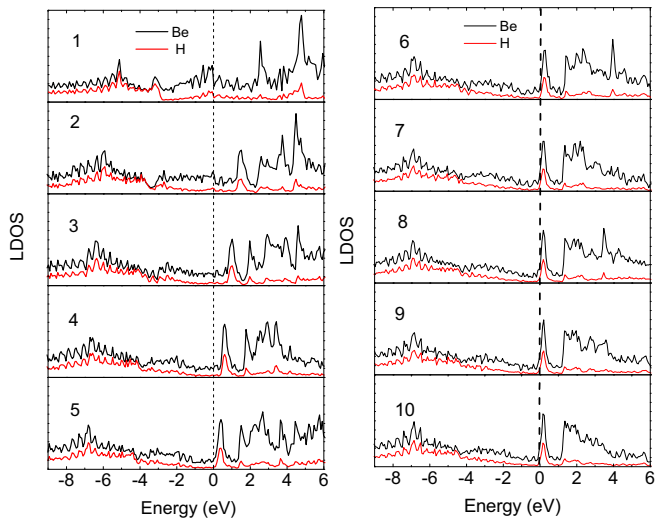}
\end{center}
\caption
{(Color online) Calculated local density of states of H adatom and topmost
Be atom. The number of Be(0001) monolayers in the slabs is shown in the figure.}
\label{fig8}
\end{figure}%
The oscillatory thickness dependence of the binding energy should be related
with the bonding properties between H adatom and surface Be atoms. To obtain a
deeper understanding of the physics behind this observation, we turn to a full
study of the surface electronic structure of H-covered Be(0001). For this we
calculate the local density of states (LDOS) through projections of the total
wave function onto atoms of interest within the Wigner-Seitz spheres around
them. The Wigner-Seitz radii for the H and Be atoms are taken to be 0.5
\AA and 1.0 \AA , respectively. Figure 8 plots a series of LDOS of H adatom
layer and the topmost Be(0001) layer for different thickness of Be thin films.
A strong chemical bonding of characteristic molecular H-Be bond can be seen.
In the case of ultra-thin slab with only one Be monolayer contained, one can
see that the Fermi level falls in the broad-peak region of Be surface. Since
the Fermi level for the clean Be(0001) surface also falls in this broad-peak
location (see Fig. 2), thus it suggests that adsorption of atomic H onto Be
monolayer does not change the surface DOS much. As a consequence, the H
binding energy is lowest in the case that only one Be monolayer is included in
the slab (see Fig. 7). When the Be(0001) monolayers in the slab is increased,
one can see from Fig. 8 that the Fermi energy continuously shifts upwards,
approaching a narrow peak, showing a more strong hybridization between H $1s$
and Be $2p$ orbitals. Consequently, the binding energy of H adatom first
curves up by increasing the Be layers in the slab, as shown in Fig. 7. When
the thickness of slab increases up to 5 Be monolayers, then the Fermi level
remains almost unchanged with respect to the narrow peak, the resultant
hybridization between H $1s$ and Be $2p$ orbitals obtains its largest
amplitude. Therefore, when further increasing the layers of Be in slab,
instead of a monotonic increase or decrease, the H adatom binding energy is
now featured by an oscillatory behavior due to metallic nature of Be(0001)
surface. Furthermore, by a comparison of the clean-surface DOS (Fig. 2) with
the H-adsorbed surface DOS (Fig. 8), remarkably, one can see that there is a
prominent change in the Be(0001) surface DOS by the atomic hydrogen
adsorption. Since it has been speculated that the metallic enhancement of Be
surface is responsible for many extraordinary phenomena such as enhanced
electron-phonon coupling\cite{Plu}, the 400 increase in the superconducting
temperature for amorphous Be compared with crystalline Be, clearly it is of
high interest to study the effect of H adsorption on these peculiar
surface-related phenomena.

\section{Conclusion}

In summary, the Be(0001) thin films have been studied by density-functional
theory pseudopotential plane-wave calculations. The dependence of electronic
structure, energetics, and interlayer relaxation upon the thickness of the
film has been fully investigated, showing the metallic QSE of the film,
although the bulk Be is nearly semiconducting. We have shown that this QSE is
associated with an interplay of surface enhancement at the Fermi level and the
bulk band splitting into subbands due to the quasi-2D electronic confinement.
We have also studied the atomic hydrogen adsorption onto the Be(0001) films.
The results gave a clear preference for bridge site adsorption. We have shown
that (i) the H adsorption energy oscillates with increasing the Be monolayers
in the slab; (ii) In the case of ultra-thin films, the H adsorbate changes the
surface density of states prominently by continuously shifting the Fermi level
upwards, approaching the narrow peak in the otherwise clean-surface DOS. The
shift will saturate after the thickness of Be(0001) film increases up to 10
layers. We expect this prominent change in the density of states will have a
significant effect on the other relevant physical properties in Be(0001) surface.

\begin{acknowledgments}
This work was partially supported by the CNSF (Grant No. 10544004 and 10604010).
\end{acknowledgments}

\end{document}